\documentclass[aps,prx,twocolumn,amsmath,amssymb,showpacs,superscriptaddress,notitlepage]{revtex4-1}
\usepackage{amsmath,amssymb,amsfonts,bm}
\usepackage{graphicx}
\usepackage{dcolumn}
\usepackage{mathrsfs}
\usepackage{bbold}
\usepackage{dsfont}
\usepackage{dcolumn}
\usepackage{epstopdf}
\usepackage[colorlinks=true,linkcolor=blue,citecolor=blue, urlcolor=blue,bookmarks=false]{hyperref}
\usepackage{changes}
\usepackage{textgreek}

\begin{document}

\title{Evaluation of real-space second Chern number using the kernel polynomial method}

\author{Rui Chen}\email[]{chenr@hubu.edu.cn}
\affiliation{Department of Physics, Hubei University, Wuhan 430062, China}

\author{Bin Zhou}\email[]{binzhou@hubu.edu.cn}
\affiliation{Department of Physics, Hubei University, Wuhan 430062, China}
\affiliation{Key Laboratory of Intelligent Sensing System and Security of Ministry of Education,\\ Hubei University, Wuhan 430062, China}
\affiliation{Wuhan Institute of Quantum Technology, Wuhan 430206, China}

\begin{abstract}
We evaluate the real-space second Chern number of four-dimensional Chern insulators using the kernel polynomial method. Our calculations are performed  on a four-dimensional system with $30^4$ sites, and the numerical results agree well with theoretical expectations. Moreover, we show that the method is capable of capturing the disorder effects. This is evidenced by the phase diagram obtained for disordered systems, which agrees well with predictions from the self-consistent Born approximation. Furthermore, we extend the method to six dimensions and perform an exploratory real-space calculation of the third Chern number. Although finite-size effects prevent full quantization, the numerical results show qualitative agreement with theoretical expectations.  The study represents a step forward in the real-space characterization of higher-dimensional topological phases.
\end{abstract}
\maketitle

\section{Introduction}
The $2n$-dimensional ($2n$D) Chern insulators are topological phases of matter characterized by quantized topological invariants known as $n$-th Chern numbers~\cite{Qi08PRB,Bernevig2013TI}. Over the past decades, much attention has been devoted to two-dimensional Chern insulators characterized by the first Chern number (also referred to as the Chern number in most contexts)~\cite{Hasan2010RMP,Qi2011RMP,Shen2017TI,Vorobev2024PRB}. In recent years, driven by breakthroughs in large-scale experimental techniques, higher-dimensional topological insulators have drawn extensive attention~\cite{Halperin87jjap,Montambaux90prb,Kohmoto92prb,
Koshino01prl,Bernevig07prl,Stormer86prl,Cooper89prl,
Hannahs89prl,Hill98prb,Masuda16sa,Tang19nat,Qin20prl,LiH20prl,
ChengSG20prb,Wang20prbrc,
Jin18prbrc,ChenR21PRL,WangCM17prl,Koh2024NC}, such as four-dimensional (4D) Chern insulators associated with the second Chern number~\cite{Mochol2018QST,Sugawa2018Science,Zhang2022CPB}, and even six-dimensional (6D) systems characterized by the third Chern number~\cite{Petrides2018PRB,Yamamoto2022PRB}. Specifically, the 4D Chern insulators have been experimentally observed in optical lattices~\cite{PricePRL2015}, photonic system~\cite{Zilberberg2018Nature}, acoustic lattice~\cite{ChenZG21PRX}, electric circuits~\cite{WangY2020NC,Yu2020NSR}.

Traditionally, Chern numbers are defined in momentum space and rely on translational invariance~\cite{Qi08PRB}. While, many realistic systems lack perfect periodicity, such as disorder~\cite{LiJ2009PRL} and quasiperiodicity~\cite{Bandres2016PRX}. To address this limitation, real-space formulations of Chern numbers have been developed and are widely used to characterize topological phases in systems without translational symmetry~\cite{Prodan10PRL,Loring2010EPL,Bianco11PRB}.  Typically, the real-space Chern number is computed via direct matrix diagonalization, which restricts the system size to approximately $10^4$ degrees of freedom~\cite{Bianco11PRB,Assuncao2024PRB}. An alternative and more efficient approach is the scattering matrix method, which extends this limit to around $10^5$ degrees of freedom~\cite{Fulga2014PRB}. Later, the kernel polynomial method has been introduced to further enhance the ability to handle large systems, enabling simulations with up to $10^7$ degrees of freedom~\cite{Varjas20PRR,Carcia15PRL,Romeral2025PRB}. Very recently, by combining kernel polynomial method with tensor network techniques, this limit can be pushed even further, reaching up to $10^9$ degrees of freedom~\cite{antao2025arXiv}.
Despite extensive investigations of the real-space first Chern number, the generalization to higher-order Chern numbers remains underexplored.

In our previous work, we used a supercell approximation to compute the real-space second Chern number of a disordered 4D Chern insulator described by the Wilson-Dirac model, achieving a maximum system size of $4^4$~\cite{ChenR2023PRB4DTAI}. However, this approach is not a strictly real-space formulation, as it relies on constructing an enlarged periodic supercell and still involves momentum-space components. Very recently, Shiina \emph{et al.} extended Kitaev's real-space formula~\cite{Kitaev2006AP} into higher dimensions, and evaluated the real-space second Chern number in the 4D Wilson-Dirac model, reaching a system size of $8^4$~\cite{Shiina2025PRB}. Although this size may appear modest, it represents a significant advancement given the explosive increase in Hilbert space dimension in higher dimensional systems. Due to finite-size effects, the quantization of the second Chern number in their results was not fully achieved~\cite{Shiina2025PRB}.

Building on these developments, we demonstrate that the kernel polynomial method enables real-space calculations of second Chern number for significantly larger system sizes, up to $30^4$. Moreover, we present the phase diagram of the system with disorder and find that the results are consistent with those obtained from the self-consistent Born approximation. Furthermore, we make the first attempt to calculate the real-space third Chern number in six dimensions. However, the numerically calculated real-space third Chern number exhibits only qualitative agreement with the expected theoretical value due to the significant finite-size effect.

\section{Model and Method}
\subsection{Wilson-Dirac model}
We start from the $2n$D Wilson-Dirac lattice model~\cite{Qi08PRB,Mochol2018QST,Ezawa19PRB},
\begin{equation}
H_{2n}(\mathbf{k})=\sum_{\mu=0}^{2n} \psi^{\dagger}(\mathbf{k}) d^\mu \Gamma_\mu^{2n} \psi(\mathbf{k}),
\label{Eq:Hamiltonian_k}
\end{equation}
where $\Gamma_\mu^{2n}$ are the gamma matrices satisfying the Clifford algebra $\left\{\Gamma_\mu^{2n}, \Gamma_\nu^{2n}\right\}=2 \delta_{\mu \nu} I$~\cite{Pais1962JMP}, with $\mu=0,1,2, \cdots, 2 n$. The components have the form:
\begin{align}
d^0=m+c \sum_{j=1}^{2 n} \cos k_j,
d^j=\lambda \sin k_j,
\end{align}
with the Dirac mass $m$, the hopping amplitude $t$ and the spin-orbital interaction $\lambda$.  Specifically, we take the gamma matrices $\Gamma^{2}=(\sigma_z,\sigma_x,\sigma_y)$ for $n=1$, $\Gamma^{4}=\left(\sigma_x \sigma_0, \sigma_y \sigma_0, \sigma_z \sigma_x, \sigma_z \sigma_y, \sigma_z \sigma_z\right)$ for $n=2$, and $\Gamma^{6}=\left(\sigma_x \sigma_x\sigma_x, \sigma_x \sigma_x\sigma_y, \sigma_x \sigma_x\sigma_z, \sigma_x \sigma_y\sigma_0,\sigma_x \sigma_z\sigma_0,\sigma_y \sigma_0\sigma_0,\sigma_z \sigma_0\sigma_0\right)$ for $n=3$, respectively. In the numerical calculations, we discretize the effective Hamiltonian on
a $2n$D hypercubic lattice and set the lattice constants as $a=1$. The volume of the system is $V=L^{2n}$, where $L=la$ are the side lengths of the hypercube. Here, $l$ is the number of sites along one direction and $l^{2n}$ corresponds to the number of total lattice sites. In the following calculations, we fix $t=\lambda=1$ and the Fermi energy $E_F=0$.

\subsection{Real-space Chern number}
We adopt the following formula to calculate the real-space $n$-th Chern number~\cite{Shiina2025PRB,Sykes2021PRB,Hannukainen2022PRL,Prodan2013JPMT},
\begin{equation}
C_n=\frac{\left(2 \pi i\right)^n}{n!} \epsilon^{j_1 j_2 \cdots j_{2n}} \operatorname{Tr}  P X_{j_1} P X_{j_2} \cdots P X_{{j_{2n}} }P,
\label{Eq:Chern}
\end{equation}
where repeated indices imply summation and
\begin{equation}
P=\sum_{E_m<E_F}\left|m\right> \left< m\right|,
\end{equation}
is the projector of the occupied bands. $X_{j_k}$ is the coordinate operator along the $j_k$-th direction. $\epsilon^{j_1 j_2 \cdots j_{2n}}$ corresponds to the antisymmetric Levi-Civita tensor. The $n$-th Chern number formula is a special case of a more general universal topological marker~\cite{ChenW23PRB}, which applies to topological insulators and superconductors in any dimension and symmetry class. For examples, for $n=1$, it reproduces the local topological marker for the first Chern number~\cite{Bianco11PRB}, with
\begin{equation}
C_1=2 \pi i \operatorname{Tr}\left(P X_1 P X_2 P-P X_2 P X_1 P\right).
\end{equation}
For $n=2$, we obtain:
\begin{equation}
C_2=-2\pi^2 \epsilon^{j_1 j_2 j_3 j_4} \operatorname{Tr}  P X_{j_1} P X_{j_2} P X_{j_3}P X_{j_4} P.
\label{Eq:C2}
\end{equation}
For $n=3$, the expression becomes:
\begin{equation}
C_3=-\frac{4\pi^3 i}{3} \epsilon^{j_1 j_2 \cdots j_5 j_6} \operatorname{Tr}  P X_{j_1} P X_{j_2} \cdots PX_{j_5} PX_{j_6} P.
\label{Eq:C3}
\end{equation}
\subsection{Kernel polynomial method}
The kernel polynomial method~\cite{Varjas20PRR,Weibe06RMP} provides a stable and efficient method to expand the action of any function of an operator
that depends on the Hamiltonian. Specifically, the projector operator is approximated by the step function~\cite{Varjas20PRR}
\begin{align}
P(\tilde{H}) &=\sum_{\chi=0}^{M} g_{\chi} \mu_{\chi} T_{\chi}(\tilde{H}).
\end{align}
Here, $M=256$ is the order of the kernel polynomial expansion, which determines the accuracy of the approximation and controls the truncation of the series. The coefficients $\mu_{\chi}$, dubbed the moments in the context of the kernel polynomial method expansions, take the form
\begin{equation}
\mu_{\chi}(E_F)=\left\{\begin{array}{ll}
1-\frac{\arccos (E_F)}{\pi} & \chi=0 \\
\frac{-2 \sin [\chi \arccos (E_F)]}{m \pi} & \chi \neq 0
\end{array}\right. .
\end{equation}
$g_{\chi}$ are the Jackson kernel coefficients~\cite{Weibe06RMP}, with
\begin{equation}
g_{m}=\frac{(M-\chi+1) \cos \frac{\pi \chi}{M+1}+\sin \frac{\pi \chi}{M+1} \cot \frac{\pi}{M+1}}{M+1}.
\end{equation}
$\tilde{H}$ is the rescaled Hamiltonian, with its energy spectrum confined to the interval $\left[-1,1\right]$.  This can be achieved by rescaling the Hamiltonian with
$
\tilde{H}=\frac{2}{\Delta E}\left(H-\frac{E^{+}+E^{-}}{2}\right),
$
where $E^{+}$ and $E^{-}$ represent the upper and lower bounds of the spectrum, respectively, and $\Delta E=E^{+}-E^{-}$. To estimate the real-space Chern number, we use the stochastic trace approximation~\cite{Weibe06RMP}:
\begin{align}
\operatorname{Tr}_S&(P X_{j_1} P X_{j_2} \cdots P X_{{j_{2n}} }P) \nonumber
\\
&\approx \frac{1}{R|S|} \sum_{i=1}^R\left\langle r_i\right| P X_{j_1} P X_{j_2} \cdots P X_{{j_{2n}} }P \left|r_i\right\rangle,
\end{align}
where $\left|r_i\right\rangle$ are random-phase vectors localized in the internal region $S$~\cite{Varjas20PRR} and $R=5$ is the number of random vectors. The projector act on a vector is calculated by
\begin{align}
P(\tilde{H})|v\rangle &=\sum_{\chi=0}^{M} g_{\chi} \mu_{\chi} T_{\chi}(\tilde{H})|v\rangle \nonumber
\\
&=\sum_{\chi=0}^{M} g_{\chi} \mu_{\chi}\left|v_{\chi}\right\rangle,
\end{align}
where $\left|v_{\chi}\right\rangle$ satisfy the recursion relation
\begin{align}
\left|v_{0}\right\rangle &=|v\rangle, \nonumber \\
\left|v_{1}\right\rangle &=\tilde{H}\left|v_{0}\right\rangle,\nonumber  \\
\left|v_{\chi+1}\right\rangle &=2 \tilde{H}\left|v_{\chi}\right\rangle-\left|v_{\chi-1}\right\rangle.
\end{align}
By applying the kernel polynomial method and the expanded projector, we can compute the real-space Chern numbers in Eq.~\eqref{Eq:Chern}.  It is noticed that the matrix diagonalization is not required with this approach, making it an efficient method for large system sizes.

\section{Second Chern number in 4D}
\begin{figure}[tpb]
\centering
\includegraphics[width=1\columnwidth]{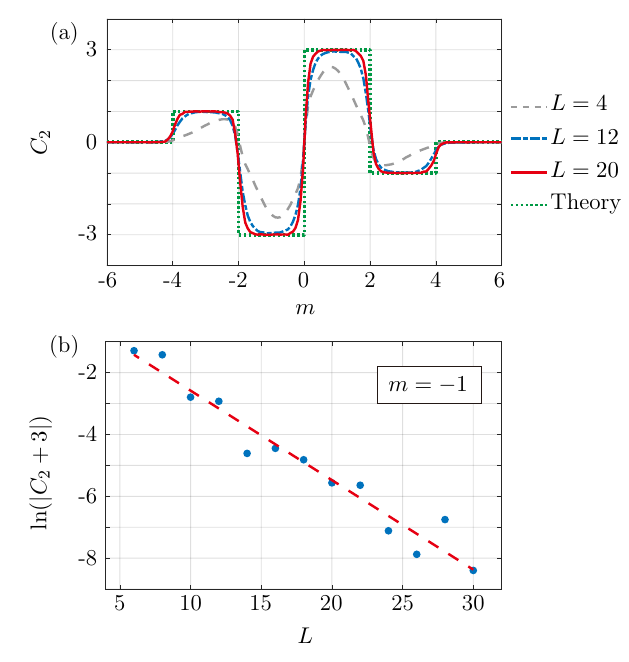}
\caption{(a) Numerically computed real-space second Chern number $C_2$ as a function of mass parameter $m$ with with different side lengths $L=4$ (gray), $L=12$ (blue), and $L=20$ (red), respectively. The green curve represents the theoretical prediction obtained from momentum-space calculations. (b) Logarithmic plot of the absolute deviation from the quantized value, $\ln\left(\left|C_2+3\right|\right)$, as a function of system size for $m=-1$. The red dashed line indicates a linear fit, demonstrating exponential convergence of the real-space Chern number toward the quantized value $C_2=-3$.}
\label{fig_4D_clean}
\end{figure}

In this section, we adopt the kernel polynomial method to calcuate the second Chern number $C_2$ [Eq.~\eqref{Eq:C2}] in the 4D Chern insulators. We begin by validating the quantization of the second Chern number in the clean case. We then show that the method is applicable to disordered systems.

\subsection{Clean Limit}
To verify the quantization of the second Chern number in our model, we compute $C_2$ numerically in real space for various system sizes. Figure~\ref{fig_4D_clean}(a) shows $C_2$ as a function of the mass parameter $m $, for different side lengths with $ L = 4, 12, 20 $. As the system size increases, the numerically obtained values gradually converge toward the theoretical prediction derived from momentum-space calculations (green curve)~\cite{Ezawa19PRB}, indicating that finite-size effects become negligible in larger systems.

To quantitatively analyze the convergence, we investigate the deviation from the quantized value $ C_2 = -3$ when $m=-1$, and plot $ \ln\left( \left| C_2 + 3 \right| \right) $
as a function of $ L $, as shown in Fig.~\ref{fig_4D_clean}(b). The linear behavior in this semi-logarithmic plot indicates that the deviation decays exponentially with the increasing system size. This exponential convergence confirms the robustness and accuracy of the real-space method for evaluating the second Chern number in finite systems.

\begin{figure}[tpb]
\centering
\includegraphics[width=1\columnwidth]{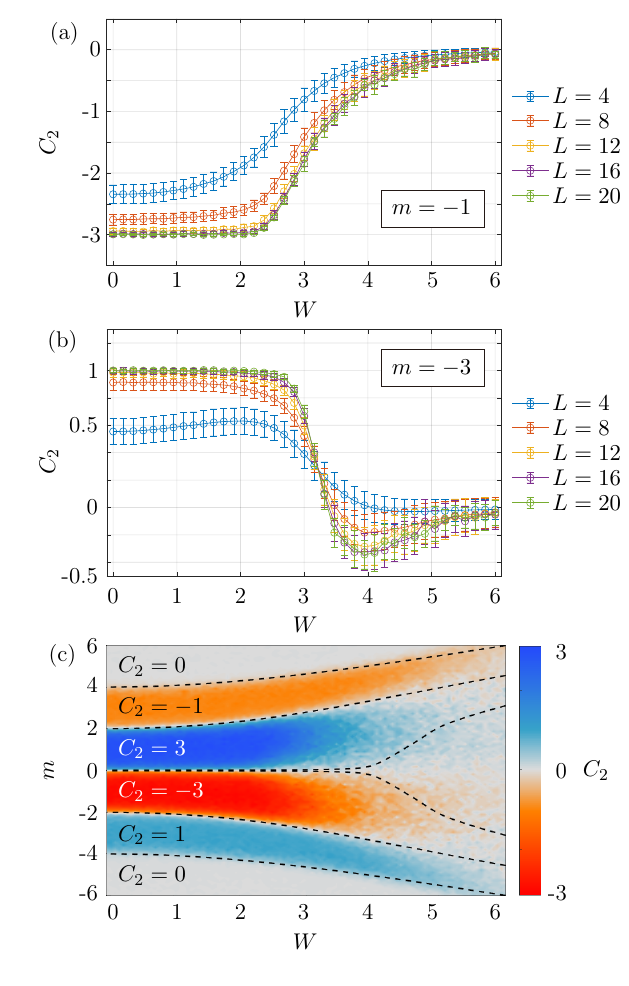}
\caption{(a)-(b) Numerically calculated real-space second Chern number as a function of disorder strength $W$ for different side length $L$ with (a) $m=-1$ and (b) $m=-3$. Here, the error bar indicates the standard deviation for 200 samples. (c) The disorder averaged $C_2$ as functions of $W$ and $m$ with $L=12$. Here, the black dashed lines are obtained by Born approximation in Appendix~\ref{Sec:Born}.}
\label{fig_4D_disorder}
\end{figure}

\subsection{Disorder effect}
We adopt the Anderson-type disorder by considering random on-site energies fluctuating with
\begin{equation}
\Delta H=\sum_{\mathbf{i}  } W_{i } \sigma_0\sigma_0 c_{\mathbf{i} }^{\dagger} c_{\mathbf{i}},
\end{equation}
where $W_{\mathbf{i}}$ is uniformly distributed within in the energy interval $[-W,W]$ and $W$ depicts the disorder strength.

Figures~\ref{fig_4D_disorder}(a) and \ref{fig_4D_disorder}(b) show the numerically computed real-space second Chern number $C_2$ as a function of disorder strength $W$ for different system sizes. The quantized value of $C_2$ remains stable at weak disorder, indicating the robustness of the topological phase under moderate disorder. As the disorder strength increases, $C_2$ exhibits fluctuations and finally collapses for strong disorder. This is in accordance with our previous works by using the super-cell approximation~\cite{ChenR2023PRB4DTAI}. Furthermore, this behavior is consistent with the findings of Ref.~\cite{ChenW24PRB}, which demonstrated that when impurities only affect the nonzero matrix elements of the lattice Hamiltonian, the average topological marker remains quantized under weak disorder.

Figure~\ref{fig_4D_disorder}(c) shows the disorder-averaged $C_2$ as a function of $W$ and $m$. The black dashed lines represent the predictions from the self-consistent Born approximation (see Appendix~\ref{Sec:Born}). For weak disorder, the numerical results are in good agreement with the Born approximation, demonstrating the validity of the kernel polynomial method for the real-space second Chern number in disordered systems. Besides, near the critical points of the homogeneous system, where the bulk gap is small, even weak disorder can drive the average topological marker away from quantized values. In contrast, far from the critical points with a larger gap, the system remains more robust and requires stronger disorder to induce deviations. This trend, also observed in lower-dimensional systems~\cite{ChenW24PRB}, indicates a generic feature that naturally extends to higher dimensions.

\section{Third Chern number in 6D}
\begin{figure}[tpb]
\centering
\includegraphics[width=1\columnwidth]{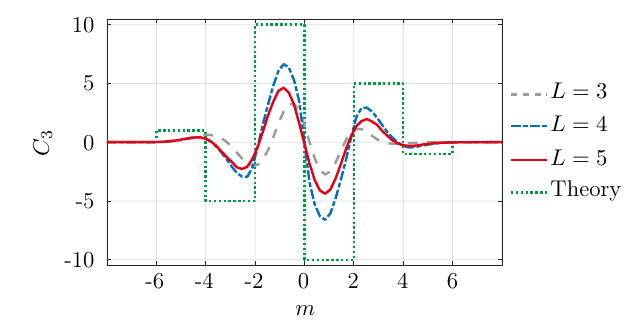}
\caption{(a) Numerically computed real-space third Chern number $C_3$ as a function of mass parameter $m$ with with different side lengths $L=3$ (gray), $L=4$ (blue), and $L=5$ (red), respectively. The green curve represents the theoretical prediction obtained from momentum-space calculations.}
\label{fig_6D_clean}
\end{figure}

Now we perform an exploratory calculation of the third Chern number $C_3$ [Eq.~\eqref{Eq:C3}] in the six-dimensional Chern insulators using the same method. As shown in Fig.~\ref{fig_6D_clean}, the numerically obtained real-space values of $C_3$ show good agreement with the theoretical predictions from momentum-space topological analysis~\cite{Ezawa19PRB}. The positions of the phase transitions are well captured, and the overall shape of the curve is consistent across different system sizes.

However, we emphasize that the values of $C_3$ are not yet quantized, and this deviation is primarily due to strong finite-size effects. Unlike the second Chern number $C_2$, where convergence can be achieved with the increasing system size. The computation of $C_3$ in real space requires significantly higher numerical complexity. This is not only due to the larger matrix dimensions in higher-dimensional systems, but also because of the involvement of the fully antisymmetric Levi-Civita tensor. In the case of $C_2$, the 4D Levi-Civita symbol $\epsilon^{j_1\cdots j_4}$ has only 24 non-zero components. In contrast, the 6D version $\epsilon^{j_1\cdots j_6}$ required for $C_3$ contains 720 non-zero terms, which drastically increases the demand for computational resources.

\section{Conclusion}

Our results demonstrate the viability of real-space approaches to second Chern numbers in moderately large 4D Chern insulators by using kernel polynomial method. However, extending the calculation of the third Chern number $C_3$ in 6D Chern insulators remains challenging due to the rapidly growing computational cost. We believe that tensor network methods~\cite{antao2025arXiv}, which can efficiently handle larger systems, hold strong potential in the future. By adopting tensor network techniques, it is possible to access significantly larger 6D real-space systems and ultimately obtain quantized values of $C_3$.

\begin{acknowledgments}
R.C. acknowledges the support of the NSFC (under Grant No.~12304195), the Chutian Scholars Program in Hubei Province, the Hubei Provincial Natural Science Foundation (Grant No. 2025AFA081), and the original seed program of Hubei University.  B.Z. was supported by the NSFC (under Grant No. 12074107), the Wuhan city key R\&D program (under Grant No. 2025050602030069), the program of outstanding young and middle-aged scientific and technological innovation team of colleges and universities in Hubei Province (under Grant No. T2020001) and the innovation group project of the natural science foundation of Hubei Province of China (under Grant No. 2022CFA012).
\end{acknowledgments}

\appendix

\section{Born approximation}
\label{Sec:Born}
To corroborate the physical interpretation of numerical simulation, we analyze
the present model within an effective medium theory based on the Born
approximation in which high-order scattering processes are neglected~\cite{Groth2009PRL}. In the self-consistent Born approximation, the self-energy
$\Sigma$ for a finite disorder strength is given by the following integral
equation%
\begin{equation}
\Sigma=\frac{W^{2}}{3}\left(  \frac{a}{2\pi}\right)  ^{4}\int_{\text{FBZ}}%
d\mathbf{k}\frac{1}{E_{F}-H_4\left(  \mathbf{k}\right)  -\Sigma
}\text{,}%
\label{Born}
\end{equation}
where $H_4\left(  \mathbf{k}\right)  $ is the 4D momentum-space Hamiltonian in Eq.~\eqref{Eq:Hamiltonian_k}. The coefficient 1/3 originates from the variance $\left\langle W^{2}\right\rangle
=W^{2}/3$ of a random variable uniformly distributed in the range $\left[
-W,W\right] $. This integration is
over the first Brillouin zone (FBZ). We use the lowest-order Born
approximation, which means setting $\Sigma=0$ on the right hand side of Eq.~(\ref{Born}).

By computing the gap-closing points of the renormalized Hamiltonian $H_4\left(  \mathbf{k}\right)+\Sigma$, we can identify the phase transition points.
The results based on the Born approximation fit well with the
numerical calculations for weak disorders [see Fig.~\ref{fig_4D_disorder}(c)]. This indicates that the kernel polynomial method for the real-space second Chern number captures the essential feature of disordered systems.

%
%
%
\bibliographystyle{apsrev4-1-etal-title_6authors}
\bibliography{refs-transport,refs-transport_v1}

\end{document}